\documentclass{aa}
\input psfig.sty
\usepackage{graphics}
\begin{document}
 
\thesaurus{8(09.18.1; 08.09.2; 03.09.5; 03.20.5)}
\title{High resolution near-infrared polarimetry of $\eta$ Carinae and the
Homunculus Nebula}
\titlerunning{Near-IR polarimetry of $\eta$ Carinae}
\author{J. R. Walsh\inst{1}
\and
N. Ageorges\inst{1,2}
}
\authorrunning{Walsh \& Ageorges}
\offprints{J. R. Walsh}
\institute{
Space Telescope European Co-ordinating Facility,
European Southern Observatory,
Karl-Schwarzschild Strasse 2, 
D85748 Garching bei M\"{u}nchen,
Germany.
E-mail: jwalsh@eso.org
\and
Physics Department, 
National University of Ireland - Galway, 
Galway,
Ireland. \\
E-mail: nancy@epona.physics.ucg.ie
}

\date{Received: 8 March 1999 / Accepted: 07 March 2000}
\maketitle

\begin{abstract}
High resolution near-infrared adaptive optics imaging and polarimetry 
have been obtained of the Homunculus Nebula and its 
central illuminating star $\eta$ Carinae (HD 93308). Images and maps  
of the linear polarization at a resolution of $<$0.2$''$ have been 
made in the J, H and K bands and in a narrow 2.15$\mu$m continuum 
band (K$_c$). The overall appearance of the nebula in the near-IR is 
similar in all bands and to that at V, with specific features less 
sharp to longer wavelengths. A comparison between the published HST 
WFPC2 1.042$\mu$m map and the AO J band image demonstrates that
ground-based AO resolution can approach that of HST. The large-scale 
pattern of polarization vectors is centro-symmetric demonstrating 
that single scattering dominates everywhere except perhaps in the 
central bright core. The only difference between the near-IR and 
optical appearance is a narrow linear feature at position angle 
320$^\circ$ extending  across the NW lobe of the Homunculus.
Polarization maps at K$_c$ in the near vicinity of $\eta$ Carinae 
were restored using a PSF derived from blind deconvolution. There 
is a definitely detected extension in the direction of the optically 
visible speckle knots and an estimate of 18\% for the polarization 
of one of the knots was made. This level of polarization suggests 
that the knots arise in dust+gas clouds in the near environment of 
$\eta$ Carinae, perhaps in an equatorial disc. 

The most remarkable result of the linear polarization mapping is the 
level of similarity in the spatial structure, and also in the degree 
of linear polarization, between the near-IR and optical data. 
Comparison of the polarization along the projected major axis of 
the Homunculus shows values in the SE lobe to within a few percent 
from V band to 2.2$\mu$m. In the NW lobe the near-IR linear 
polarization values agree to within a few percent over the 
1.2-2.2$\mu$m range, but are upto 10\% lower than at V.
Such a polarization pattern cannot arise in Mie scattering from 
a single power law distribution of grain sizes unless the particles 
are very small compared with the wavelength, in
disagreement with mid-IR observations. In addition the colour
dependence of the extinction was found to be shallower than the 
typical ISM, indicating the presence of large grains. 
Several possibilities are explored to try to explain these
contradictory results. Optical depth effects and a broad 
distribution in grain sizes are favoured, perhaps with a 
dependence of the grain size with depth into the small obscuring 
clouds in the lobes of the Homunculus. However the presence of 
aligned grains, previously inferred from mid-IR polarization, may 
also affect the scattered radiation from the dust.

\end{abstract}

\section{Introduction}
$\eta$ Carinae,
situated in the Carina complex at about 2.3kpc (Davidson \& Humphreys 
\cite{dahu97}), is 
one of the most massive stars known in the Galaxy and is going through
the Luminous Blue Variable phase (Humphreys \& Davidson \cite{huda94})
of unsteady mass loss (Davidson et al.
\cite{da86}). During the 1840's it underwent an outburst and reached
visual magnitude -1; since then is has been emerging from the dust 
which condensed after this ejection (Walborn \& Liller, \cite{wali}). 
Long term monitoring of optical, IR, radio and X-ray spectra has 
revealed evidence of periodicity perhaps related to a binary or multiple
star at the core of the nebula (Daminelli et al. \cite{dam97}).

The compact nebula around $\eta$ Carinae (HD 93308), called the Homunculus,
was first shown by Thackeray (\cite{tha56}) to be highly polarized. 
The initial polarimetry was 
confirmed by Wesselink (\cite{wess}) who measured linear polarization 
of around 40\%. Visvanathan (\cite{vis})
observed that the polarization centred on $\eta$ Carinae was almost
constant with wavelength from U to R and increased with increasing aperture 
size. In a small aperture, higher polarization was observed on the NW
side of the nebula than on the SE. The first systematic
polarization maps were made by Warren-Smith et al. (\cite{warr}) in the
V band and demonstrated a centro-symmetric pattern of polarization
vectors with a marked asymmetry in the polarization values along the
major axis (position angle $\sim$130$^\circ$) with values upto
40\% in the NW lobe. To produce such high
values of polarization in a reflection nebula, Mie scattering by 
silicate particles with a size distribution weighted to smaller particles
was invoked and modelled by Carty et al. (\cite{cart}). In the near
($\leq$0.5$''$) vicinity of $\eta$ Car itself, speckle masking 
polarimetry in the H$\alpha$ line and local continuum has revealed 
evidence for a compact equatorial disc aligned with the minor axis 
of the Homunculus (Falcke et al. \cite{falc}). Within $<$1$''$ of $\eta$ 
Car the polarization vector pattern does not remain centrosymmetric
in the R band, suggesting that local structures and perhaps 
intrinsic emission may contribute to the
morphology and scattered light (Falcke et al. \cite{falc}). Polarimetry
in the mid-infrared, where the dust emits rather than scatters radiation, 
shows an entirely different pattern of polarization vectors with a trend
to be oriented radially, particularly
near the boundary of the emission (Aitken et al. \cite{ait95}).
Such a pattern can be interpreted in terms of emission from aligned 
grains; Aitken et al. (\cite{ait95}) suggest that the alignment 
mechanism may be gas-grain streaming, driven by the high outflow velocity, 
or a remnant magnetic field from a dense magnetized disc.

There is a wealth of IR observations of $\eta$ Car and the Homunculus
on account of its intrinsic IR brightness, first observed by
Westphal \& Neugebauer (\cite{wene}), and astrophysical interest.
The IR spectrum is characterized by a peak around 10$\mu$m, 
indicative of silicate grains (Mitchell \& Robinson, \cite{miro}).
There is a central IR point source together with a second peak on the minor
axis of the nebula, whose separation increases from 1.1 to 2.2$''$ 
from 3.6 to 11.2$\mu$m (Hyland et al. \cite{hyl}). The near-IR spectrum
of $\eta$ Car shows a steep increase with wavelength and 
prominent hydrogen lines of the Paschen and Brackett series as well
as He~I lines (Whitelock et al. \cite{whi}) and weaker Fe~II and [Fe~II]
lines (Altamore et al. \cite{alta}). Maps in the J, H and K 
bands show that the structure is dominated by scattering, but beyond 
about 3$\mu$m dust emission dominates (Allen \cite{all89}), with 
many clumps present. High spatial resolution
observations have reported an unresolved central source
(at L and M band, Bensammar et al. \cite{bens}), with filaments and
unresolved knots within 1$''$ detected in many IR bands (Gehring
\cite{gehr}). Maps in the mid-IR show a similar structure and the
compact central source has a dust temperature $\sim$650K and dust mass
of 10$^{-4}$M$_\odot$ with nearby dusty clouds associated into
loop features (Smith et al. \cite{smai}). 
This source has been so prodigiously studied at so many
wavelengths that it possesses its own review article in Annual 
Reviews of Astronomy and Astrophysics (Davidson \& Humphries 
\cite{dahu97}).

  $\eta$ Car can be considered an ideal source for adaptive optics on 
account of its very bright central, almost point, source 
(V$\sim$7mag. - van Genderen et al. \cite{vang}) and the limited radial 
extent ($\pm\leq$10$''$) of the Homunculus, which means that the source 
itself can be used as a reference star for the wavefront sensor. As a
consequence, off-axis anisoplaniticity does not significantly 
affect the adaptive optics (AO) correction out to the
edges of the nebula. Previous near-infrared AO imaging of $\eta$ Car 
was obtained (Rigaut \& Gehring \cite{rige}), including some limited 
polarimetry (Gehring \cite{gehr}) using
the COME-ON AO instrument. Here we report on dedicated high
resolution near-IR AO imaging polarimetry conducted at J, H, K, and 
in a continuum band at 2.15$\mu$m, using the
ADONIS system and SHARP~II camera with the aim of studying the 
small-scale polarization structure of the Homunculus.
The observations are described in Sect. 2; the reductions 
and polarization data are presented in Sect. 3
and the relevance of the results for the structure and dust
properties of this remarkable reflection nebula are discussed
in Sect. 4.

\section{Observations}
Imaging polarimetry of $\eta$ Carinae was obtained with the 
ADONIS Adaptive Optics instrument mounted at the F/8.1
Cassegrain focus of the ESO 3.6m telescope. ADONIS is 
the ESO common-user adaptive optics instrument; it employs a
64 element deformable mirror and wave-front sensor (Beuzit et al. 
\cite{bez97}).
For the observations of $\eta$ Car, a Reticon detector was
used as the wave-front sensor. The camera is SHARP~II -
a Rockwell 256$^{2}$ HgCdTe NICMOS 3 array (Hofmann et al. \cite{hof95}).
The pixel scale was chosen as 0.050$''^{2}$, giving a field of view of 
12.8$\times$12.8$''$. Although this field does not encompass
the full extent of the Homunculus nebula it was chosen so
that well-sampled diffraction limited imaging would be possible
in at least the H and K bands. Table 1 lists the observations;
the two orientations, referred to as A and B, had $\eta$ Car 
in the lower right and upper left of the array respectively
(east is up; north to the right) enabling full coverage of the 
Homunculus.
The narrow band 2.15 $\mu$m (henceforth K$_c$) observations were 
made with $\eta$ Car in the centre of the array. For each filter 
and orientation, exposures were made 
at 8 position angles of the polarizer: 0.0, 22.5, 45.0, 67.5,
90.0, 112.5, 135.0, 157.5$^\circ$ in sequence. In
addition a repeat of the 0$^\circ$ exposure was made with the
polarizer angle set to 180$^\circ$ in order
to check the photometry and repeatability of the polarimetric
measurements. Each full sequence of polarization measurements at
the 8+1 position angles was repeated as specified in column 
6 of Table 1. Offset sky chopping was employed and the relative
position of the offset sky is listed in column 5 of Table 1; the
exposure on sky was equal to the on-source time. As discussed
in Ageorges \& Walsh (\cite{agwa}) it was not possible to calculate
reliable K band polarimetry; these data will therefore only be 
discussed in terms of their high resolution imaging.
On 1996 March 03 due to a technical problem, the computer control
of the chopper malfunctioned and the offset sky had to be observed 
subsequent to each sequence of polarizer angles, and in some cases 
the exposure time on source was greater than on background sky.
The last column of Table 1 gives an indication of the external
seeing as measured by the Differential Image Motion Monitor at 
La Silla during the observations. Photometric standards were not 
observed and no attempt has been made to determine accurate
magnitudes in the J, H, K and K$_c$ filters.

\begin{table*}
\caption[]{ADONIS Polarimetry Observations of $\eta$ Carinae}
\begin{flushleft}
\begin{tabular}{lrrrrrrr}
Field & Filter~~~ & No.~~ & Exp. & Offset sky & No.~~ & Date~~~~ & Seeing \\  
  & $\lambda_{c}$,$\Delta \lambda$ ($\mu$m) & Frames & (ms) & $\Delta \alpha$,$\Delta \delta$($''$) & Repeats & & ($''$)~~ \\
\hline
KA    & 2.16,0.16 & 200~~ & 50~ & 0,-18~~~~ & 2~~~~ & 1996 Mar 01 & 1.2~~ \\
KB    &           &     &    & -18,0~~~~ & 2~~~~ &            &  \\
HA    & 1.64,0.18 & 200~~ & 50~ & 0,-18~~~~ & 2~~~~ & 1996 Mar 02  & 0.8~~ \\
HB    &           &     &    & -18,0~~~~ & 2~~~~ &         &  \\
JA    & 1.25,0.15 & 200~~ & 50~ & 0,-18~~~~ & 1~~~~ & 1996 Mar 03 & 0.7~~ \\
JB    &           &     &    & -18,0~~~~ & 1~~~~ &     & \\
K$_c$    & 2.145,0.017 & 100~~ & 50~ & -18,0~~~~ & 2~~~~ & 1996 Mar 03 & 0.8~~ \\
\hline
\end{tabular}
\end{flushleft}
\end{table*}

Polarized and unpolarized standards were observed in the course of
the observations to determine the instrumental polarization and
any rotation of the instrumental plane of polarization. The star 
HD~64299 which is relatively close to $\eta$ Car was observed at J, 
H and K as an unpolarized standard (polarization 0.15\% at B (Turnshek
et al. \cite{turn}) and
assumed to be low in the IR, although the actual values are not measured)
and a point source for deconvolution. In the K$_c$ 
filter, the star HD~94510 was observed. These stars were chosen 
primarily to be bright but not so bright as to saturate the 
SHARP~II camera. ADONIS observations of these
stars and the polarized reference sources are
fully described in Ageorges \& Walsh (\cite{agwa}), where the
observational details are also given.

\section{Reductions and Results}

\subsection{Basic reduction}

The data cubes at each position of the polarizer consists of
M $\times$ 256$\times$256 pixel frames, where M is given as the number of frames in
Table 1. To produce a single image for each of the nine selected
polarizer angles, the data were flat fielded using images obtained at the
beginning of the night on the twilight sky with an identical set of
polarizer angles. The data cubes were used to derive a bad pixel map
as described by Ageorges \& Walsh (\cite{agwa}) using a
sky variation method. The sky from the offset position was subtracted 
separately from each of the M frames before combination into
a single image for each polarizer angle.
These reductions were performed with the dedicated adaptive optics
reduction package `eclipse' (Devillard \cite{devil}). The data frames 
at position angle 157.5$^\circ$ were not used in any computations
of polarization on account of the discrepancy noted by 
Ageorges \& Walsh (\cite{agwa}). The reduced data then consisted of
2 sets of K band image pairs; 2 sets of H band image pairs;
and 1 set of J band images; all with $\eta$ Car displaced 
from the centre of the detector and 1 set of K$_c$  filter images
centred on  $\eta$ Car.

\subsection{Registration of images}

The rotation of the polarizer induces a small
shift of upto 3 pixels in the position of the images (see
Ageorges \& Walsh \cite{agwa}, Fig. 2) and coupled with
the (intended) shifts of $\eta$ Car across the SHARP~II field, it
is necessary to carefully align all images to a common centre in order
to calculate precise colour or polarization maps for the whole of
the Homunculus. On the J, H and K frames, the image of $\eta$ Car
was saturated (overflow of A-to-D converter). The centroids of
all the J, H and K images were determined using a large radius
(30 pixels = 1.5$''$); it was found that with such a large radius 
the centroid was not sensitive to the saturated core (typically 
a few pixel radius). Combined images of the coverage of the whole nebula,
at each position of the polarizer, were formed by shifting, and 
rotating by 90$^\circ$, each image pair (e.g. JA and JB which were 
taken consecutively) 
to a common centre and averaging the pixels in common. The rotation was
required to produce astronomical orientation. Shifts were 
restricted to integer pixels, thus the alignment can have a maximum 
error of 0.050$''$. The alignment procedure was carefully checked
by examining the coincidence of features in the nebula and in
the core when saturation was not severe (e.g. in the J band images in
particular). Where there were repeats of the full combined
image (column 6 of Table 1), the two sets of images were
averaged. The result was an image of dimensions 326$\times$326
pixels (16.3$\times$16.3$''$) with no data in the top left and 
lower right corners. Figure 1 shows the J, H, K and K$_c$ total
flux images (i.e. Stokes I) on a logarithmic scale. All images
have identical scale and orientation. In the J, H and K images
the saturation of the central source is indicated by the zero flagged
pixels (region of radius 4 pixels about position of peak). There are 
a variety of artifacts: diffraction spikes along the principal
axes caused by the secondary support; low level changes consequent
on merging images (the overlap regions were used to scale the
image pairs); a doughnut shaped feature caused by a hot pixel cluster 
which occurs at equal declination values at the
extremity of the NW and SE lobes and near the rim of the NW lobe.
These features, which are most apparent on the colour maps (Fig. 2),
have not been masked out but are obviously not interpreted.
 
\begin{figure*}
\centerline{\vbox{
\hspace{1.0cm}
\psfig{figure=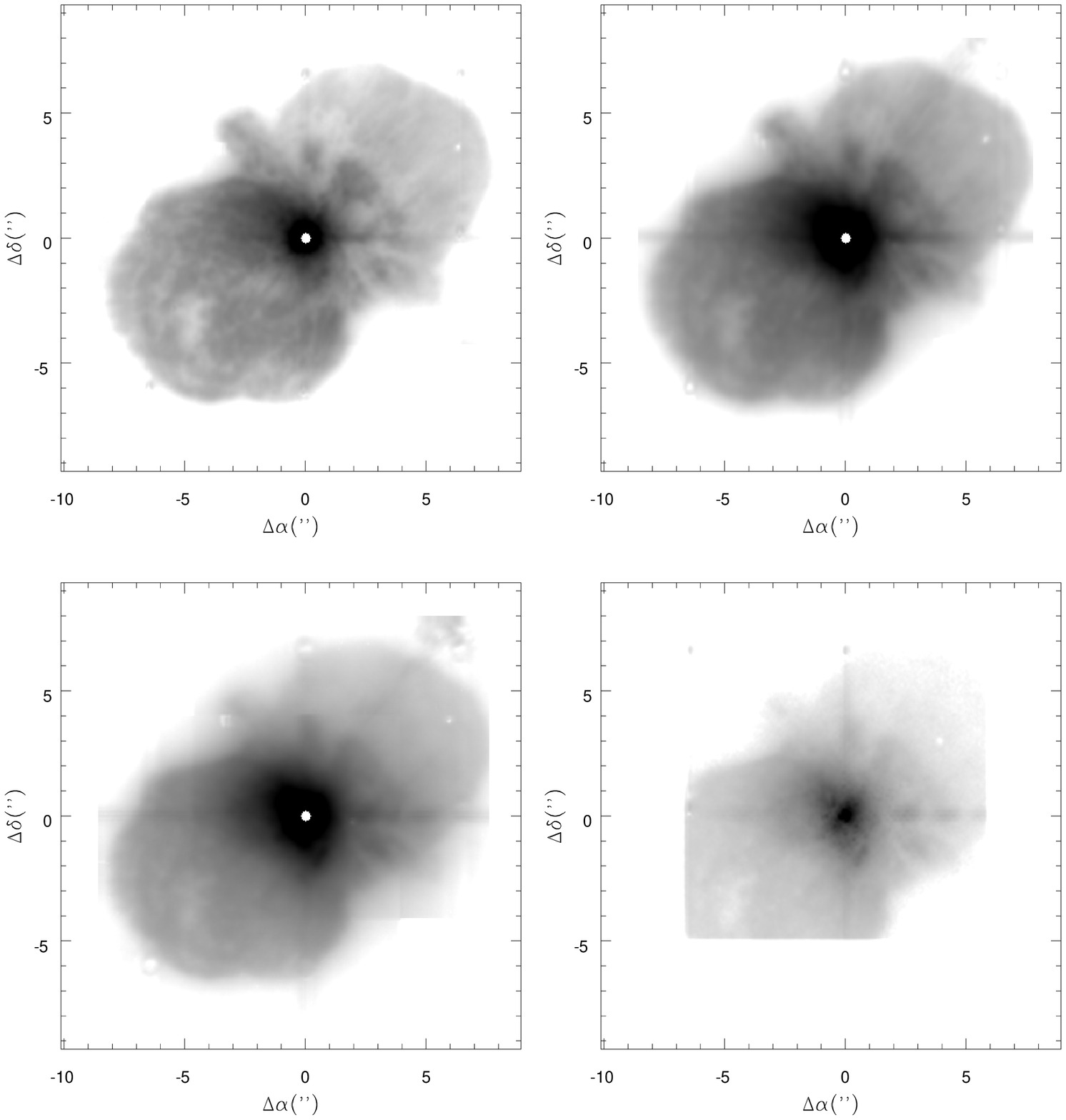,height=18cm,angle=0,clip=}
}}
%\resizebox{19.0cm}{!}{\includegraphics{rub1.ps}}
\caption{The final combined J (upper left), H (upper right),
K (lower left) and K$_c$ (lower right) total intensity (Stokes I)
images of the Homunculus 
nebula are shown in logarithmic plots. The dynamic range in the plots 
is about 10$^{4}$ and north is at the top; east to the left. The 
4 pixel radius region about the position of $\eta$ Carinae containing
saturated values has been set to zero in the J, H and K maps. The
axes display the offset positions from $\eta$ Carinae.}
\end{figure*}

\subsection{Colour maps} 
`Colour' maps were made by ratioing the J, H and K images. A cut-off
in the form of a mask was applied to each colour map in order to
prevent division by small numbers and produces the sharp bulbous 
edges in the maps. Figure 2 shows the J/H and H/K images on a logarithmic 
scale. On account of saturation the values over the core do not
hold any colour information and have been set to zero. The range of 
valid ratio values are: (J-H) 0.02 - 2.5 mag.; (H-K) 0.02 - 1.5 mag.

\begin{figure*}
\centerline{\vbox{
\hspace{0.0cm}
\psfig{figure=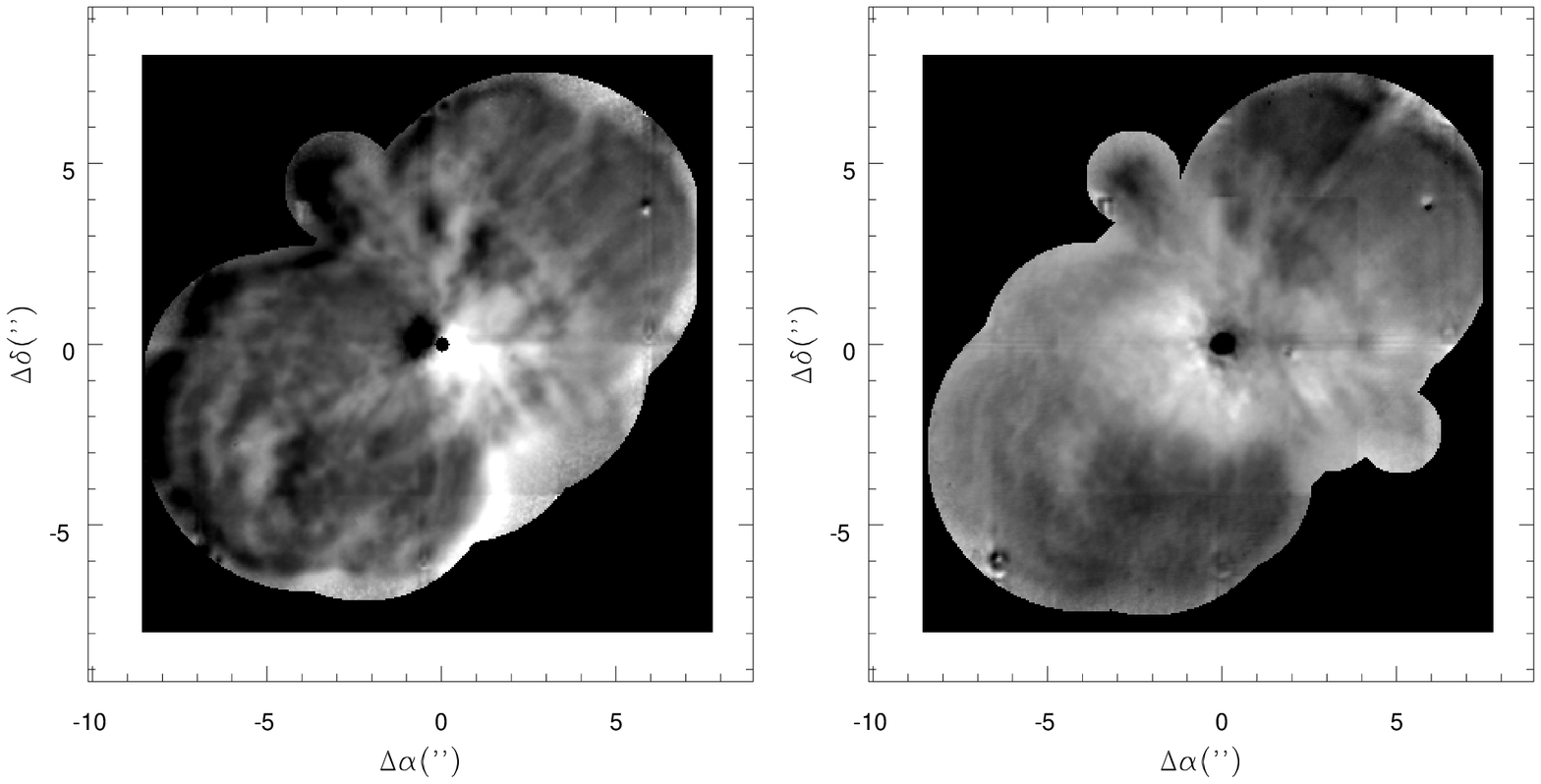,width=18cm,angle=0,clip=}
}}
%\resizebox{19.0cm}{!}{\includegraphics{etacarfig2.ps}}
\caption{The logarithmic colour maps of J/H (left) and H/K (right)
are shown for the Homunculus nebula. The plotted range is
2.5 mag. for J-H and and 1.5 for H-K. Masks have been applied
to suppress the outer low signal level and the saturated core. 
The size and orientation are identical to Fig. 1.}
\end{figure*}

\subsection{Polarization maps}
The linear polarization and position angle were calculated for the combined
maps by fitting a cosine 2$\times \theta$ curve to values at each point
as a function of polarizer angle as fully described in 
Ageorges \& Walsh (\cite{agwa}). The discrepant point at PA 157.5 
was not included in these fits (Ageorges \& Walsh \cite{agwa}). 
The input images were binned to improve the signal-to-noise in the
polarization determination at the expense of spatial resolution. In 
addition a cut-off in polarization signal-to-noise (i.e. $p / \sigma_p$) 
was applied to exclude points with large errors, 
such as at the edges of the Homunculus. 
Figure 3 shows the J, H and K$_c$ band polarization vector maps 
superposed on logarithmic intensity contour maps to be directly
compared to the images in Fig. 1. The data
were binned into 4$\times$4 pixels (0.2$\times$0.2$''$) before
calculating the polarization; the polarization cut-off was set
at errors of 2\% for the J and K$_c$ maps and 1.7\% for the H band map. 

\begin{figure*}
\centerline{\vbox{
\hspace{0.5cm}
\psfig{figure=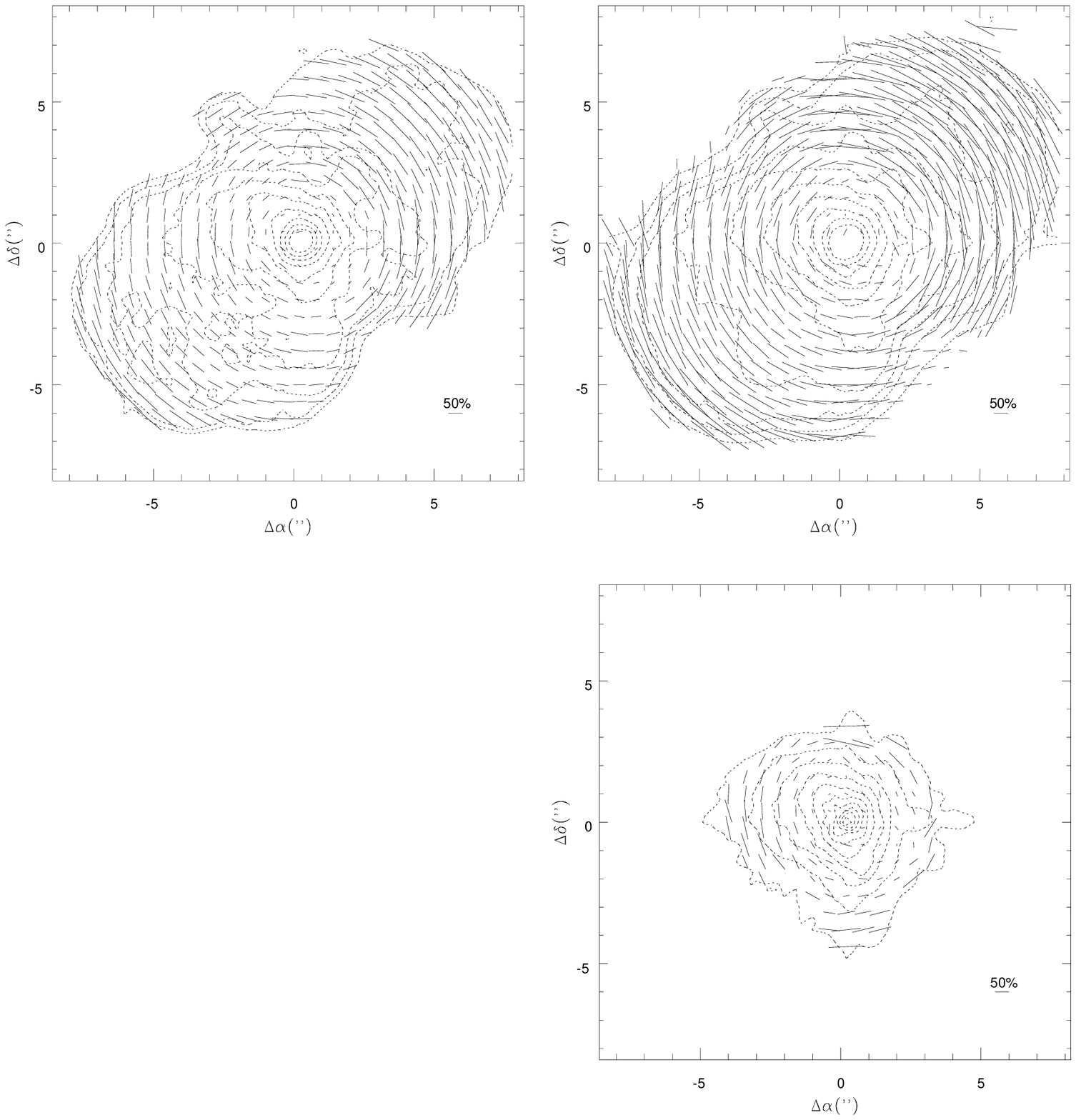,height=18cm,angle=0,clip=}
}}
\caption{The polarization vector maps of the combined J (upper left), 
H (upper right) and K$_c$ (lower right) images of the Homunculus 
nebula are shown together with logarithmic contour of the total counts. 
The orientation is as Fig. 1 and the size of the polarization vectors
are indicated in \%.}
\end{figure*}

\subsection{Restoration of 2.15$\mu$m images}

As described in Ageorges \& Walsh (\cite{agwa}) the K$_c$ images
were restored using the blind deconvolution algorithm IDAC
(Jeffries \& Christou \cite{jeff}) to determine the PSF of 
the images. Only the central 128$\times$128
pixel area was restored to save computer time and only the 
images at polarizer angles of 0, 45, 90 and 135$^\circ$ were
employed. Figure 4 shows the logarithmic total intensity map 
over the central 1$\times$1$''$ area resulting from 
restoring the four reduced images using the Richardson-Lucy
algorithm (Lucy \cite{luc74}) with the IDAC PSF; the resulting
image was reconvolved with a Gaussian of 3 pixel FWHM (0.15$''$)
since this is about the expected diffraction limited resolution 
at this wavelength. 

  A polarization map was calculated from the four restored 
1$\times$1$''$ images and is shown in Fig. 4 for direct comparison 
with the logarithmic image. The data were binned 2$\times$2 pixels 
before calculating polarization and the polarization
cut-off error was 4\%. An attempt was made to calculate the
polarization at the positions of the speckle knots, discovered by
Weigelt et al. (Weigelt \& Ebersberger \cite{wei86} and Hofmann 
\& Weigelt \cite{hof88}), whose positions are shown in Fig. 4.
Knot A is the central (assumed) point source whilst knots B, C and D 
are to the NW at offsets of 0.114 (B), 0.177 (C) and 0.211$''$ (D); 
these offsets correspond to only 2.3, 3.5 and 4.2 pixels in the
images. In the restored images no distinct knots could
be discerned at these positions but it is clear from Fig. 4
that there is an apron of IR radiation in the NW direction
strongly hinting on an area of elevated brightness in the vicinity
of these knots. 

Aperture photometry of the Weigelt et al. knots in a 2$\times$2 pixel 
area was performed  for the three sets of images -  
restored with IDAC, Richardson-Lucy restored with the IDAC PSF and
Richardson-Lucy restored. All the restorations were convolved with a 
Gaussian of 3 pixel FWHM. For the three images the aperture polarization
determinations showed  that knot B could not be distinguished from knot
A (identical polarization within errors). Knot C showed very differing 
results depending on the method (it lies on a diffraction spike); 
only for knot D could a fairly consistent value of polarization be 
determined. From the three methods a mean polarization of 18 $\pm$ 7 \% 
and a position angle of 17 $\pm$ 14 $^\circ$ was derived for knot D. 
Given the position angle of knot D from knot A of 336$^\circ$ (Hofmann 
\& Weigelt \cite{hof88}), a position angle of the polarization vector 
of about 60$^\circ$ is expected. To reconcile this discrepancy,
it is suggested that knot D may not be directly illuminated by 
$\eta$ Carinae, i.e. there is multiple scattering 
within this core region which would not be too surprizing given
the high (gas) densities (Davidson et al. \cite{davebb97}). The 
mean total intensity
ratio knot A/knot D was 10:1, to be compared with the value of about
12:1 given by Hofmann \& Weigelt (\cite{hof88}) for a wavelength of
$\sim$8500\AA. It is justified to attempt polarimetry at these 
positions since the images of Morse et al. (\cite{morse}, Fig. 5) 
and Ebbets et al. (\cite{ebb94}) show no obvious
indication that the knots have substantial proper motion.
This supposition is partly supported by the low radial velocities
measured by Davidson et al. (\cite{davebb97}) who classify these 
knots as `compact slow' ejecta from $\eta$ Car.

\begin{figure}
\vskip 0.0cm
\centerline{\hbox{
\psfig{figure=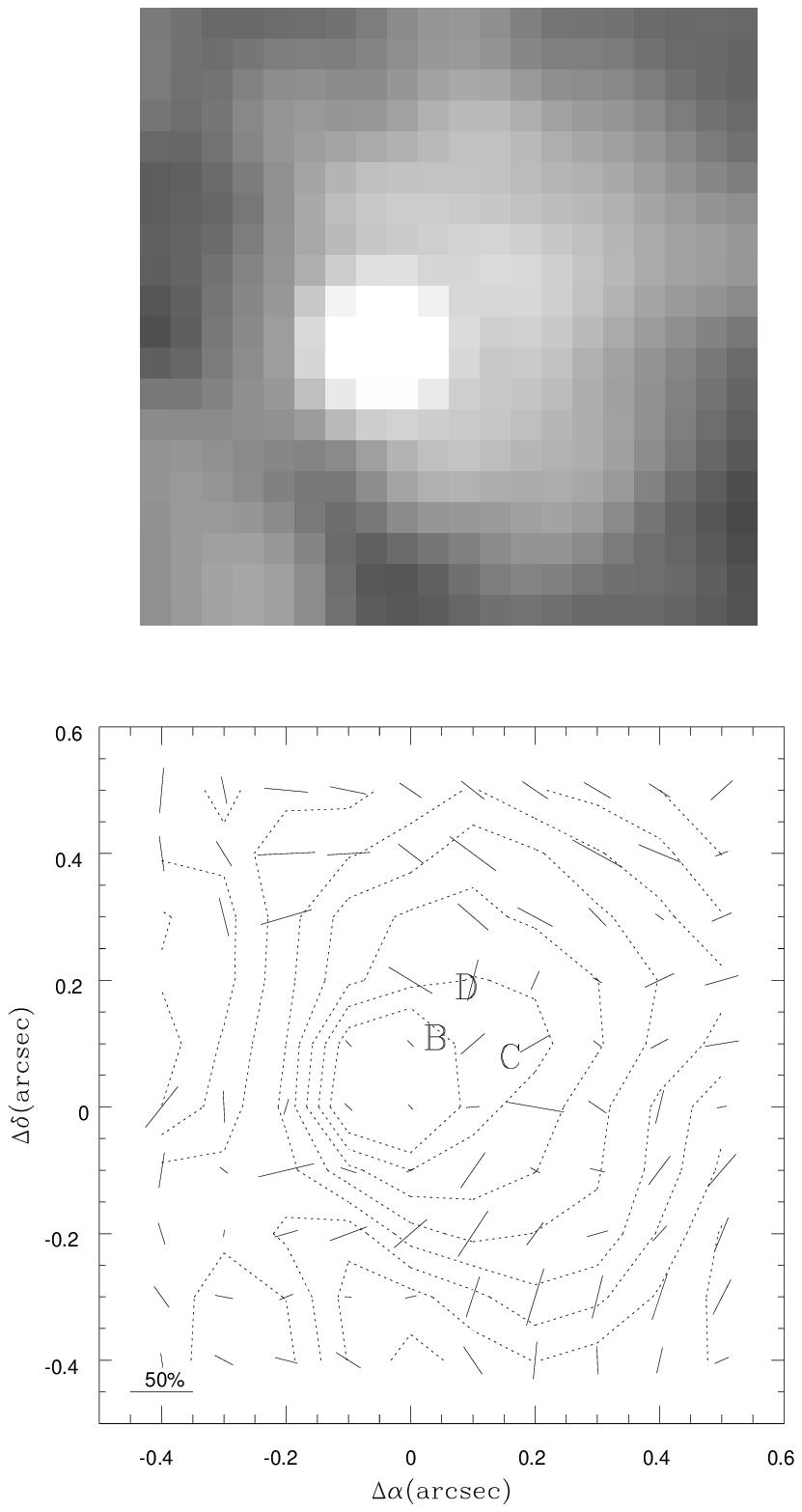,height=16.0cm,angle=0,clip=}
}}
\caption{A logarithmic restored K$_c$ image of the
1$\times$1$''$ region centred on $\eta$ Carinae is shown 
in the upper plot. The image was restored with the
Richardson-Lucy algorithm using the PSF from the IDAC
deconvolution and the result was convolved with a Gaussian 
of 3 pixels width. The lower image shows the
polarization vector map of the same region superimposed on a 
logarithmic contour map of the total count. The polarization 
vectors were calculated for 2$\times$2 pixels binned. The positions 
of the Hofmann \& Weigelt (\cite{hof88}) knots B, C and D are 
indicated on the vector map. 
}
\end{figure}

\section{Discussion}

\subsection{Morphology}
In the near-IR, scattering still dominates the structure of 
the Homunculus nebula as it does in the optical, and the general
appearance is similar. The features in the colour maps in Fig. 2
correspond to those well known in the Homunculus 
(see for example the sketch of the various morphological features 
in Fig. 3 of Currie et al. \cite{curr}).
The paddle to the NW, which is bluer, as are the two knots at 
PA 0 and 280$^\circ$, are interpreted as belonging to the
disk in which $\eta$ Carinae resides (see e.g. the sketch
of Smith et al. \cite{smigeh}, Fig. 10). The jet NN to the 
NE is weaker at H and is barely detected at K, probably on account 
of lower extinction towards this jet in comparison to the Homunculus 
(assuming that scattering dominates this structure). 
The skirt to the SW is less striking in all the IR maps
compared with the optical (see the beautiful HST images
reproduced in Morse et al. \cite{morse}) and does not show
a distinct colour in Fig. 2. The SE lobe presents a more speckled 
appearance than the NW one where there are some radial features which
show up well in the colour maps. The most prominent large  scale
feature in the H-K colour map (Fig. 2) is the dark trench
extending over most of the NW lobe at PA $\sim$320$^\circ$.
This feature is barely visible in the J and H maps but is
much brighter at K. The `hole' in the SE lobe, detected 
in the mid-IR by Smith et al. (\cite{smigeh}), has rather 
red J-H and H-K colours; its
western edge is noticeably brighter at J. The hole in the
NW lobe, detected in the mid-IR images of Smith et al. (\cite{smigeh}),
is not visible in the near-IR images. The rim of the
SE lobe is notably blue in the J-H map whilst it is barely
discernable in the H-K map; the rim of the NW lobe is notably redder. 
These differences must be primarily due
to differing amounts of line of sight extinction at the periphery of 
the lobe:- the SE lobe which is tilted toward the observer 
suffers less extinction than the NW lobe which is tilted away.

The trend noted by Morse et al.
(\cite{morse}) that structures become less pronounced 
with increasing wavelength continues into the near-IR. 
Figure 5 (upper) shows a cut in Log flux along the major axis 
(taken as PA 132$^\circ$) of the Homunculus from the 
J, H, K and K$_c$ maps shown in Fig. 1. The central 2$''$  
is not shown for the J, H and K maps since the images of
$\eta$ Carinae are saturated in this region. The effect of
smoothing out of features is clear from this plot. This is
more strikingly seen in Fig. 6 where an attempt has been
made to display the near-IR flux distribution (linear scale)
along the same cut shown by Morse et al. (\cite{morse})
[their Fig. 6]. Note in particular the depth of the 
feature centred at offset $+$1.3$''$ which shows a central depression
of 40\% of the peak value at $+$1.9$''$ for a wavelength of 1.25$\mu$m, 
compared to 94\% at 0.63$\mu$m. 
The lower contrast with increasing wavelength can be attributed to lower
extinction, through the Homunculus lobes, of many small ($^{<}_{\sim}$0.3$''$)
optically thick knots. Such knots block the transmission of
scattered radiation from $\eta$ Carinae through the front side of 
the lobes and on account of their optical thickness do not 
show much scattered light from their near sides.
This can account for the more dappled appearance of
the nebula in the J-H than the H-K colour map (Fig. 2).

Figure 6 reveals that at some positions 
of the nebula there can be substantial differences in the structure
with wavelength: the peak at $+$3.3$''$ in the K band, which is hardly 
noticeable at J, is the most prominent feature in this comparison. This
peak is seen on the colour maps in Fig. 2 as the dark region in 
the H-K map south of $\eta$ Carinae. Whilst there are some
colour differences over the near-IR range towards the edge of the
nebula the most prominent are in the central ($\sim$4$\times$4$''$) area. 
The knots in the
NW lobe within 3.5$''$ of $\eta$ Carinae are stronger in the J image
than at longer wavelengths. This region is also marked out by having
a distinctly lower polarization and corresponds to the disc
(e.g. Smith et al. \cite{smigeh}), which has a very different orientation 
to the Homunculus. The axis of the Homunculus is assumed to be tilted 
at about 35$^\circ$ to the plane of the sky (e.g. Meaburn et al. 
\cite{mea93}; Davidson et al. \cite{davebb97}). However the 
HST proper motion studies favour a higher inclination of about 50$^\circ$
(Currie, 1999, priv.comm); the double flask model of Schulte-Ladbeck
et al. (\cite{schul}) has a 60$^\circ$ tilt to the plane of the sky. 
Differences in structure between the optical and
near-IR can be understood in terms of increasing optical thinness 
with wavelength; at K the outer regions of the disk are optically thin
and the sightline extends to the core of the Homunculus. This is also 
confirmed by features of the polarization maps (Sect. 4.2).

 The J band image was compared in some detail with the HST F1042M 
image presented by Morse et al. (\cite{morse}) as Fig. 4. The 1.04$\mu$m 
HST image, which differs by only 0.24$\mu$m in central wavelength from
the J band image. These images make an
excellent image pair for comparing HST with ground-based
adaptive optics, although the AO image has not been deconvolved. 
The J band image is definately `fuzzier'. This must partly
be due to the trend for features to be smoother at longer wavelengths
but probably is primarily due to the differing character of the
AO PSF compared with HST, since the
diffraction limits are comparable (0.07$''$ at J for ESO 3.6m 
and 0.09$''$ at 1.04$\mu$m for HST). Clearly the
ground-based AO image is approaching the HST image in terms of
resolving sharp features close to the diffraction limit. The only
difference noticed between the images was the presence of a narrow
bright feature running through the dark region at approximately
$\Delta \alpha$=1,$\Delta \delta$=-2$''$ on the J band image.
Presumably this distinct feature corresponds to a shaft of radiation 
escaping from the central disc.
  
The restored K$_c$ map of the central region shown in Fig. 4 
definitely shows an extension
in surface brightness in the direction of the speckle knots
(Weigelt \& Ebersberger \cite{wei86} and Hofmann \& Weigelt 
\cite{hof88}). As discussed by Ageorges \& Walsh (\cite{agwa})
several methods were used for restoring this image and all
showed the presence of this feature, so its reality seems probable.
That it is visible at K, in the UV and optical, strongly suggests
that dust scattering is the common spectral feature, although
these knots are known to have extraordinary line emission
(Davidson et al. \cite{davebb95}; Davidson et al. \cite{davebb97}; 
Davidson et al. \cite{da99}).
In Fig. 7 of Davidson et al. (\cite{davebb97}), the brightness
profile of the continuum in the NW shows an extension in comparison
with the SE direction and this was suggested as scattered light from
dust in the speckle condensations C and D. The detection of
an extension in this direction at K$_c$ (the filter avoids the
Brackett $\gamma$ and He~I lines, so is presumably pure
continuum), confirms this interpretation. The polarization
value derived for knot D in Sect. 3.5, although rather
uncertain, suggests a scattering origin. Comparing the
images in Fig. 4 with those presented in Fig. 5 of Morse
et al. (\cite{morse}), over almost identical regions, shows
similarities to the NW of the central source but 
no detailed correspondence to the SE. This can be understood
in the canonical picture that here the disk, tilted by
some 35$^\circ$ to the line of sight, is being viewed
obliquely and the radiation escapes preferentially towards the observer
to the NW. That any UV continuum is visible at all to the NW indicates
that the extinction must be fairly low and confirming the knots B-D 
as features on the nearside of the obscuring disc material. 

The only `new' morphological feature to come from the
high resolution IR maps of the Homunculus is the linear feature at
PA 320$^\circ$. This is seen on the H and K images and well seen in the H-K 
colour map (Fig. 2), but not in the J band (Fig. 1).
The position angle of the linear feature points back to the 
position of $\eta$ Carinae and coincides with that of one of
the whiskers detected outside the Homunculus by Morse et al. 
(\cite{morse}) - WSK320 (see also Weis \& Duschl \cite{weis}). 
This `whisker' has a high positive velocity (Weis \& Duschl \cite{weis}) 
and a suggested high proper motion (Morse et al. \cite{morse}).
The linear feature is also aligned with the only region of IR flux 
detected beyond the extent of the Homunculus, designated as NW-IR. 
This knot appears to be a rather diffuse cloud of  extent $\sim$2$''$, 
which is very red: the K/H flux ratio of the knot is twice that of
the nearby region of the NW lobe of the Homunculus. NW-IR is highly 
polarized at H and well detected;
the value of linear polarization is 36\% in a 0.5$''$ aperture. This
value is slightly larger than the corresponding values for the edge
of the Homunculus nearest the knot (see Fig. 3 where the feature
is apparent and Fig. 5 for the polarization at the edge of the
Homunculus). It is a real feature as it was observed
on both H and K band images (the signal-to-noise was not large enough to 
detect it on the K$_c$ image) and is presumably a dusty knot on the
symmetry axis of the NW lobe. It was not however seen on the 2.15$\mu$m
image of Smith et al. (\cite{smigeh}) perhaps on account of lower
signal-to-noise. 

If the whiskers are high velocity ejecta then considering their high  
length-to-width ratio, an extreme collimation mechanism is suggested. 
The detection of an aligned feature inside the Homunculus suggests 
that this could be a spatially continuous jet feature extending from 
close to $\eta$ Carinae. There is slightly elevated polarization (3-4\% above
the mean of the surroundings) on this narrow feature with a trend to
lower polarization on both sides of its length (1-2\% less). The elevated
polarization suggests that the feature cannot be intrinsic line emission 
which would dilute the polarized flux. From the 
HST WFPC2 images, the whiskers are however bright in [N~II] line emission.
The detectable polarization within the Homunculus and the presence of a 
dust cloud at the end of this linear feature suggests a confined dust+gas 
feature. The NW-IR dust cloud is however in strong  
contrast with the highly confined line emission. It cannot be directly
claimed that both are aspects of the same phenomenon although it is
highly suggestive. It is suggested that the reason the 
linear feature is not seen at optical wavelengths and in the J 
band is on account of its being confined to the inside of
the Homunculus where there is enough extinction to mask it
at lower wavelengths. Clearly this feature
would repay further study at high spatial resolution and with 
spectroscopy. No IR features were convincingly 
seen associated with any of the other whiskers.
%However it would be valuable to search for features within the Homunculus
%which are aligned with the whiskers. 
The jet NN is probably 
ballistic (e.g. Currie et al. \cite{curr}) and the whiskers may be also;
so detectable remnants may be expected extending back to $\eta$ Carinae 
itself. The confusion by extinction and dust scattering however makes this a 
difficult task.

\begin{figure}
\centerline{\vbox{
\psfig{figure=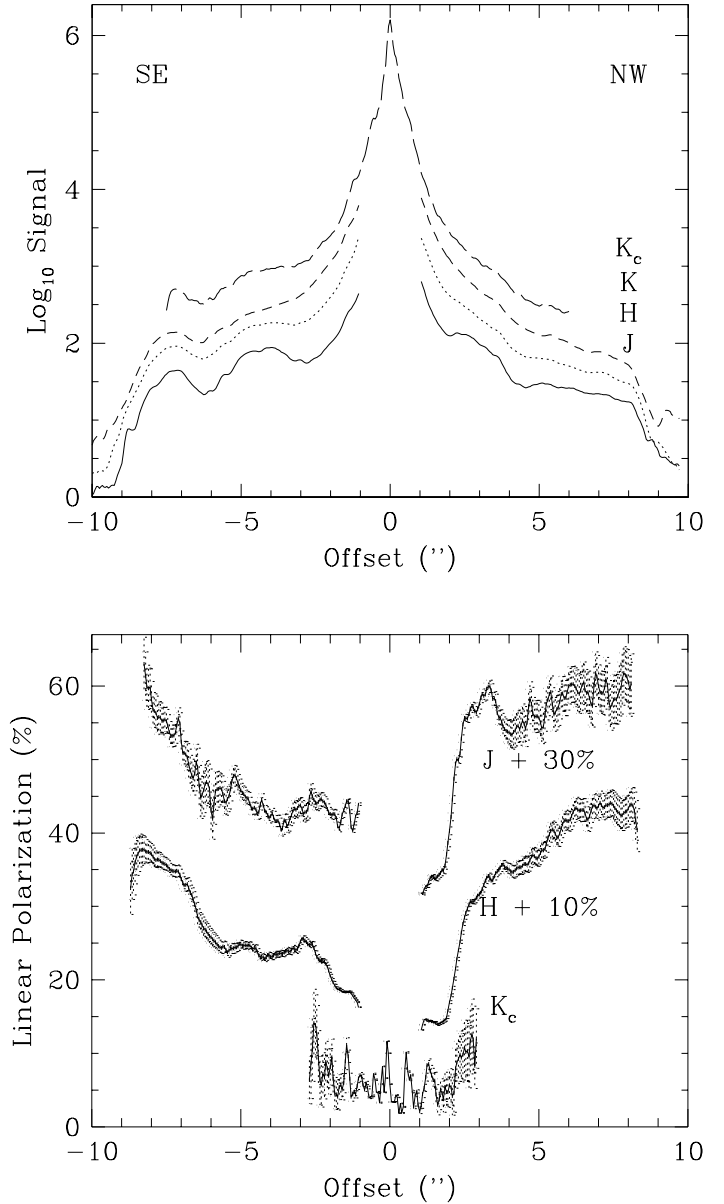,height=16cm,angle=0,clip=}
}}
\caption{A stacked plot of the logarithmic relative signal along 
the major axis of the nebula (PA 132$^\circ$) in the J, H, K and 
K$_c$ bands is shown in the upper plot. The traces have been 
vertically displaced for clarity and the saturated region of
the J, H and K images has not been plotted. The lower plot shows
the variation in linear polarization (\%) along the same
axis for the K$_c$, H and J data, with the J and H data 
vertically displaced. The errors on the polarization 
measurements are also indicated.}
\end{figure}

\begin{figure}
\centerline{
\hbox{
\psfig{figure=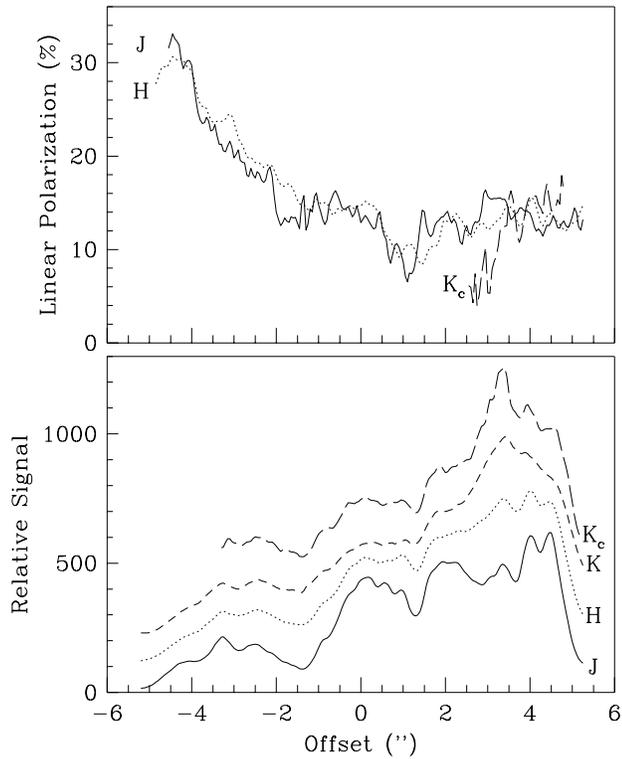,height=10cm,angle=0,clip=}
}}
%\resizebox{\hsize}{!}{\includegraphics{etacarfig2jh.ps}}
\caption{The linear signal in the J, H, K and K$_c$ images
across the SE lobe of the Homunculus corresponding to
the intensity cut of Morse et al. (\cite{morse}, Fig. 6)
is shown in the lower plot. In the upper plot the J, H and K$_c$ 
polarization (\%) is shown along the same cut. The origin has 
been set to coincide with that of the Morse et al. cut}
\end{figure}

\subsection{Linear Polarization structure}

The polarization maps shown in Fig. 3 have a smooth appearance.
The structure of the polarization vectors shows no strong evidence for 
diverging from the characteristic centro-symmetric pattern indicating 
illumination by a central source. 
This result is in contrast 
to the optical polarization maps of Warren-Smith et al. (\cite{warr}),
which show a slightly elliptical pattern of polarization vectors. 
This difference indicates that the dust must be substantially optically 
thick in the central waist in the optical but thin in the J to K region.
The overall smoothness of the polarization maps indicates that there 
cannot be substantial variations in the positions of the scattering 
centres along the line of sight through the lobes otherwise the
polarization would vary between say the back and front of the lobe 
(assuming a Mie scattering origin in which the polarization depends solely
on scattering angle). However if the nebula is composed of
optically thick small clouds then the scattering is always from the
side facing the observer and no single scattered flux is received from
the rear side of a dense cloud. The overall smoothness of the
centro-symmetric pattern also shows that there must not be much, if any,
multiple scattering occurring in the Homunculus itself.
The central region in the restored K$_c$ polarization map (Fig. 4)
appears to show some regions which depart from the centrosymmetric pattern.
However these regions coincide with the positions of the telescope spider 
where the polarization determination is unreliable.

\begin{figure}
\centerline{
\hbox{
%\hspace{1cm}
\psfig{figure=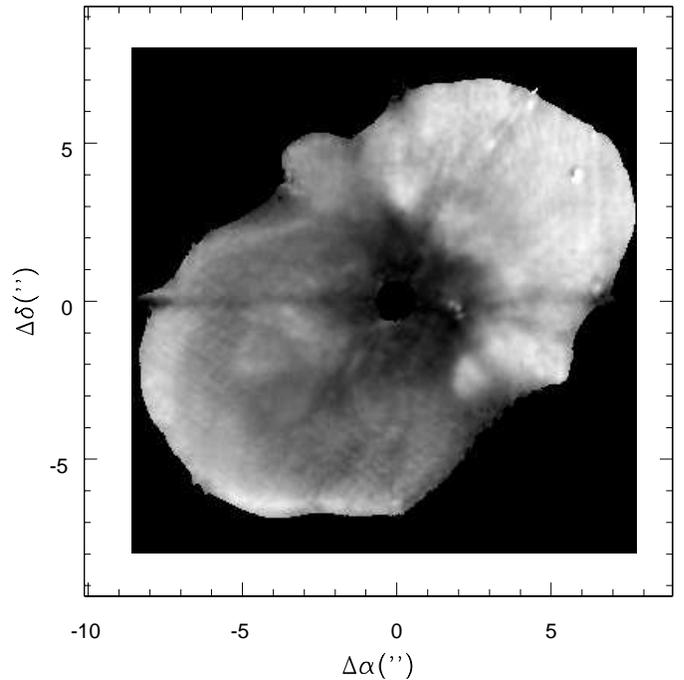,height=9.0cm,angle=0,clip=}
}}
\caption{The unbinned map of the H band degree of linear polarization  
is shown plotted on a linear scale. The plotted range is 0 (black) to 
37\% (white). The central core, which was saturated on the raw data, 
has been set to zero polarization as have the outer regions with low
signal.
}
\end{figure}

Comparison of the polarized and unpolarized
images at J and H does not show intermediate scale features ($\sim$ few times
the diffraction limit) with lower polarization, corresponding to emission 
regions. The dominant features are the lower polarization central region 
and the more highly polarized lobes. The polarization structure along
the NN jet to the NE is very similar in both J and H and shows a
plateau at 20\% at distances from 3.6 to 5.8$''$ offset
(in PA 33$^\circ$). The demarcation between the
regions of the higher polarization lobes and the lower polarization central
region is rather abrupt especially to the NW and W (the change from 
black to white on Fig. 7); to the SE it merges into the lower polarization of
the lobe tilted towards the observer. The lower polarization 
region centred on $\eta$ Carinae is roughly rectangular in shape
(see the H band polarization map - Fig. 7) with dimensions 
5.5$\times$5.0$''$; the longer axis is perpendicular to the major
axis of the Homunculus. If this region is interpreted
as the disc with an inclination $\sim$90$^\circ$ to that of the Homunculus, 
then the NW region has the smaller inclination to the line of sight
and the low polarization could arise on account of small scattering
angles. The SE section of the disc should be behind the SE lobe and would 
therefore have low polarization on account of large scattering angles, be faint
and relatively obscured by dust in the foreground lobe. For an
inclination of the disc of $\sim$35$^\circ$ to the line of sight,
the deprojected diameter is 10$''$ which is similar to the projected
minor axis diameter of the nebula (PA 42$^\circ$) excluding the
NN jet and the skirt to the SW. It has been suggested that the 
central low polarization region reflects a change in grain 
properties (Smith et al. \cite{smigeh}). However if the grains were 
smaller in the disc than in the lobes, then the polarization should be 
larger at long wavelengths.

The values of the linear polarization along the projected major axis
at J, H and K$_c$ (Fig. 5) are very strikingly similar. 
This can be compared to the polarization along the same axis 
at V shown in Warren-Smith et al. (\cite{warr}) [Fig. 2 
with direction reversed] and Schulte-Ladbeck
et al. (\cite{schul}), Fig. 9. From a low plateau of about 5\% in the 
central region, the polarization rapidly increases to about 20\% 
at 2.5$''$ NW and less steeply to 13\% at 2.5$''$ SE. 
In the SE lobe the polarization values are identical to within 2\%
at J and H (i.e. within the errors). The shape of the polarization 
profile is also identical
to that at V (Schulte-Ladbeck et al. \cite{schul}) but the V values 
are 3\% higher on average with the SE edge about 6\% higher, i.e.
definately larger than the typical errors.
In the NW lobe the J polarization has a pronounced peak at $+$3.3$''$;
this peak is also apparent at V but less pronounced   
at H. The J and H values both increase steadily from 25\% at
4$''$ offset to 33\% in J and 30\% in H band at 8$''$ offset. At V
the polarization has a different behaviour: it is higher (around 40\%), 
flatter with offset and shows a decrease beyond 7$''$.

The change in the magnitude of the peak at 3$''$ offset NW with 
wavelength can be interpreted as caused by different extinction 
optical depths in the various bands. At J (and V), the scattering arises 
predominantly in the equatorial disc, which has an inclination
of about 35$^\circ$ and thus gives rise to polarization from scattering
angles at around this value; at H the optical depth is lower and the
disc begins to become transparent at this wavelength and there is a greater
contribution of scattered flux from the rear lobe of the Homunculus.
Further out in the NW lobe the H band polarization is systematically a 
few percent higher than at J whilst the V band polarization is about 10\% 
higher. The difference between V and J band polarizations is readily 
understood in terms of the scattering at longer wavelength arising from deeper
within the lobe since the line of sight extinction optical depth is
lower. The V band polarization in the NW lobe, which is tilted away from the 
line of sight, arises predominantly from the nearside
of the lobe where the scattering angle is closer to 90$^\circ$.
The elevated polarization in the H band over the J band is
the reverse of the trend of polarization decreasing with wavelength,
but consistent with the presence of grains small with respect to the
wavelength. However the deduction from the mid-IR data and colour temperature maps 
of Smith et al. (\cite{smigeh}) suggest the hotter ($\sim$400K) core 
dust is caused by smaller ($\sim$0.2$\mu$m)
grains and the cooler ($\sim$200K) outer lobes by larger (1-2$\mu$m) 
silicate grains (see also Robinson et al. \cite{rob87}). None of these
suggested grain sizes can explain the small changes of polarization with
wavelength in the optical-IR range.

In Fig. 6 (upper) the linear polarization at J, H and K$_c$ across the
cut of Morse et al. (\cite{morse}) is shown. The similarity of overall
values is again apparent as for the projected major axis (Fig. 5), but 
there are local differences. At offset $-$1.5 to $-$2.0$''$ for example,
there is a distinct dip in the polarization at J by $\sim$4\% and at
offset $+$3 to $+$4$''$ the peak in emission at K$_c$ has lower polarization 
than at J and H. These differences can be interpreted as due to
scattering from material at different depths within the nebula
suffering differing amounts of extinction giving rise to different scattering
angles and hence lower polarization. The dip at $-$1.5 to $-$2.0$''$ is
accountable by extinction at J biasing the polarization to regions nearer
to the observer; the region at $+$3.5$''$ coming from a more extincted
rearward region, perhaps in the equatorial disc. 

\subsection{Dust properties and structure}
The most surprizing result of the IR polarization measurements is that 
the polarization along the long axis shown in Fig. 5 is so similar in J 
and H. In the SE lobe the polarization values are also within a few percent 
of those at V. This was totally unexpected. For scattering by grains
small enough to produce 30\% polarization at V, the H band
polarization should be 80-90\% since the particles are now much smaller 
than the wavelength (approaching the Rayleigh scattering regime).
Alternatively the grains are very small and Rayleigh scattering
occurs at all wavelengths; however then it is not clear why the 
polarization values are not larger. If there is a substantial
unpolarized component which dilutes the polarized flux, say from
a different grain size population, then it would be expected that
this is wavelength dependent. The only strong wavelength dependent
difference is the higher polarization at V than in J and H by about 
10\% in the NW lobe, the reverse of the behaviour expected by 
scattering of grains small with respect to the wavelength.
 
One possibility for the lower than expected polarization at J and H
from the Mie scattering prediction could be dust emission from
the warm grains. This would become more significant with longer
wavelength; thus some depolarization would be expected at K.
The similarity of the K$_c$ and H polarization shows that little
depolarization is detected, thus refuting any influence on
the J and H polarization values. If the grain properties were changing
with position in the nebula this would be expected to have an effect 
on the behaviour of polarization at different wavelengths. In the
NW lobe the differences between V and near-IR polarization can be
attributed to differing scattering angles. If the line-of-sight
optical depth is low the scattering region dominating the 
observed flux is deeper inside the lobe than for a high line-of-sight 
optical depth. The scattering angle for the lower line-of-sight
optical depth will be larger. In the SE
lobe there is no large-scale difference in polarization from V to K,
corresponding to a situation where the line-of-sight optical depth 
is low or does not vary much. 

In order to highlight this conclusion in Fig. 8 
the polarization is plotted at four
positions in the nebula defined by square 0.5$\times$0.5$''$ apertures.
These positions were chosen in regions where the polarization 
distribution is fairly flat to give a representative estimate
rather than an average over a wide range of values. The positions were 
chosen at ($\Delta \alpha$, $\Delta \delta$): ($+$4.75,$+$4.85), 
($+$1.15,$+$0.55), ($-$0.55,$-$1.70) and ($-$2.05,$+$0.55) arcsec, 
representing offset distances of 6.8, 1.3, 1.8 and 4.9$''$ from $\eta$ Carinae
respectively. Assuming that the axis of the Homunculus is tilted
by about 35$^\circ$ to the plane of the sky (Meaburn et al. \cite{mea93}
Davidson \& Humphreys \cite{dahu97}),
the scattering angles of the first and last regions are 125 and 65$^\circ$
(or 140 and 40$^\circ$ for a 50$^\circ$ tilt to the plane of the sky). 
The region at ($+$1.15,$+$0.55) arcsec is expected to be in the disc and
thus have a scattering angle about 35$^\circ$. The scattering
angle for the third region ($-$0.55,$-$1.70$''$) is not easily
predicted and the polarization was used to place it at an
appropriate scattering angle by interpolation. A value of about 
45$^\circ$ is suggested.

An attempt was made also to plot the scattered flux in Fig. 8 by scaling 
the total counts in the images to the near-IR photometry of Whitelock
et al. (\cite{whi94}). Since the near-IR magnitudes are decreasing
with time, an approximate extrapolation was made to the year of observation;
the following total magnitudes for $\eta$ Carinae and the Homunculus
were adopted: J 2.7; H 1.9; K 0.6. Comparison with the plot of J 
magnitude of $\eta$ Carinae from 1970 to 1999 in Davidson et al. 
(\cite{da99}, Fig. 3) shows the J band estimate to be reliable.
The zero points for the magnitude system were taken from Koornneef 
(\cite{koor}). No correction was attempted for the slight saturation on 
the central point source which affected the J and H images. The
K$_c$ image was used to scale the saturated K band flux. No attempt was
made to correct for line of sight extinction. The lower panel
of Fig. 8 shows the resulting fluxes in the apertures. The
scattered flux decreases with scattering angle as expected for
Mie scattering. The H and K fluxes generally agree fairly well
whilst the J band fluxes are higher; this would be consistent with
the scattering angles actually being less for the J band 
consequent on the depth at which the predominant scattering
is viewed being less on account of line of sight extinction.

\begin{table}
\caption[]{Extinction to globule at $+$1.3$''$ in Morse cut over SE lobe}
\begin{flushleft}
\begin{tabular}{ccc}
$\lambda$ ($\mu$m) & $\Delta$ mag. & $\Delta$ mag. \\
                   & Observed      & ISM $^\ast$\\
\hline
0.336  & 2.90  & 3.90 \\
0.410  & 3.25  & 3.25 \\
0.631  & 3.06  & 1.98 \\
1.042  & 1.75  & 0.91 \\
1.250  & 0.49  & 0.66 \\
1.650  & 0.23  & 0.39 \\
2.150  & 0.22  & 0.23 \\
2.200  & 0.14  & 0.22 \\
\hline
\end{tabular}
\end{flushleft}
$^\ast$ Normalised to the observed value at 0.410$\mu$m \\
\end{table}

  One way to attempt to visualize the extinction is to examine
the wavelength variation of scattered light emerging at a position where
there is an extinction feature. A position such as the pronounced 
drop in surface brightness at $+$1.3$''$ on the Morse et al. 
(\cite{morse}) cut (Fig. 6) appears to be promising. 
The fractional depth of this feature was measured relative to the
peaks by linearly interpolating between the values at $+$0.5 and $+$2.0$''$.
This is clearly dependent on the spatial resolution, especially when
it comes to estimating the peaks which are much sharper than the
extinction hole. The result is given in Table 2 expressed in
magnitudes (i.e. A$_{\lambda}$ from the WFPC2 and ADONIS 
measurements). For comparison the expected extinction for a
Galactic law matched to A$_{4100A}$ is listed in column 3
(using the Seaton (\cite{seat}) extinction law as parametrized by
Howarth \cite{how83} and R=3.1).
For the Morse et al. (\cite{morse}) data, the F336W point appears
anomalous; it may be that the peaks also suffer extinction so the
measurement of the extinction to the globule is underestimated. 
Over the range 4100 - 10420\AA\ the extinction towards the nebula drops
but less sharply than for the Galactic extinction law. There is clearly
a jump in values between the HST 1.042$\mu$m and the ADONIS J band extinctions,
probably on account of differing spatial resolution and PSF's causing 
differing degrees of infilling. Treating the measurements
from J to K separately in Table 2 shows that the extinction 
also drops less steeply than the Galactic extinction curve, strengthening
the suggestion of a flatter extinction law in this region of the
Homunculus. This somewhat greyer extinction favours particles
larger than those typically found in the ISM, in agreement with the
conclusions of Smith et al. (\cite{smigeh}) and others on dust emission.
Davidson et al. (\cite{da99}) also suggested grey extinction from a comparison 
of the modestly wavelength dependent brightening of
$\eta$ Carinae and the Homunculus in the optical and near-IR.
     
\begin{figure}
\centerline{
\hbox{
%\hspace{1cm}
\psfig{figure=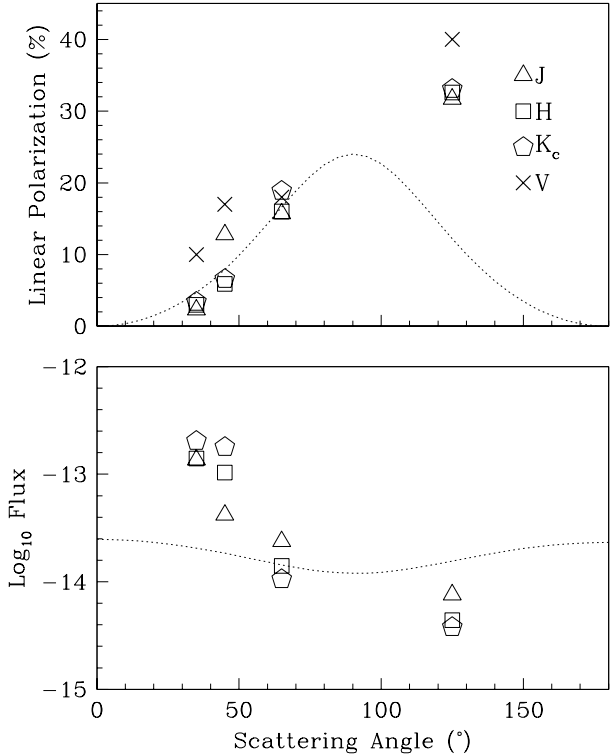,height=10cm,angle=0,clip=}
}}
\caption{The behaviour of the J, H and K$_c$ polarization (upper) 
and Log$_{10}$ flux (lower) from four selected positions in the Homunculus
plotted as a function of the assumed scattering angle 
for these positions. The positions are at ($\Delta \alpha$($''$), 
$\Delta \delta$($''$)): 
($+$4.75,$+$4.85), scattering angle 125$^\circ$; 
($+$1.15,$+$0.55), scattering angle 35$^\circ$;
($-$0.55,$-$1.70), scattering angle 45$^\circ$ (deduced); and
($-$2.05,$+$0.55), scattering angle 65$^\circ$. 
Also shown on the upper plot is the V band polarization deduced,
at similar offset positions, on the Homunculus major axis from
Schulte-Ladbeck et al. (\cite{schul}), Figs. 5 and 9. The expected
variation of polarization and scattered flux is shown by a dotted line
for Mie scattering at H band by silicate grains of size 0.065$\mu$m, the 
largest grains which can produce similar polarizations at V to K band. 
The curves have been arbitrarily normalised to the 65$^\circ$ scattering
position.
}
\end{figure}

There is a strong discrepancy between the predictions of
the extinction and emission of grains in the Homunculus and their
polarization properties. If the grains were typically 1-2$\mu$m in
the Homunculus, Mie theory for spherical particles predicts a maximum 
polarization at 1.65$\mu$m of 38\% for a scattering angle of
120$^\circ$ (assuming that the size distribution
is flat from 1-2$\mu$m and using the Draine (\cite{drain})
optical constants for silicate grains). However the V band 
polarization from Mie theory is only 13\% for such a size range of particles
at the same scattering angle. Whilst Rayleigh scattering from
very small $^<_\sim$0.1$\mu$m grains produces 
a similar polarization at all wavelengths, it produces only a small
variation in scattered flux with scattering angle (by a factor $\sim$2). 
Thus Rayleigh scattering is not capable of matching the 
points as shown by the scattered intensity and polarization
curves in Fig. 8 (see caption for details). It has not so far been
possible to find a single size distribution which would explain the 
consistent polarization value over a wide wavelength range and the similar 
variation in scattered flux with scattering angle from J-K (see Fig. 8). 
To explain the consistency of the polarization, other suggestions
involving a grain size distribution together with optical depth
effects which `tune' the scattering properties with wavelength must be
invoked. 

Three possibilities are suggested to explain the dust scattering
structure in the Homunculus: \\
a) the grains possess a range in size which is similar at all
positions within the nebula but the extinction of this grain 
distribution at a given wavelength is such that the particles which
contribute most to the scattering have lowest extinction. In other
words when the extinction cross section is low, the scattering cross
section is high. From Mie theory this is not possible for a single
grain species but could occur for some grain mixture. The extinction
acts to fine tune the size range contributing to
scattering. It is assumed here that the scattering angle changes
rather little with wavelength (hence extinction). For the 
polarization to stay constant with wavelength, there
would need to be a grain size distribution inside
small clumps. The unit optical depth scattering surface would then
be deeper at longer wavelengths on account of the lower extinction.
Grain-gas or grain-grain collisions in the high velocity clouds
could perhaps explain the size distribution which would affect the 
surface regions more strongly; \\
b) the effective scattering angle alters with wavelength on account of
the differing extinction. To longer wavelengths the small dust globules 
become more transparent, resulting in an increase in scattering angle and 
a greater proportion of the scattered flux arises from towards the rearside
of the lobes. This could compensate the increase of polarization with 
wavelength by providing less polarized flux. Whilst this could work for
the NW lobe it is not easily able to explain the polarization behaviour in
the SE lobe. Here the scattering angle increases with increasing penetration 
(lower line-of-sight extinction) into the lobe and so the polarization should be
expected to increase with increasing wavelength. In this case the relevant 
parameter is again the line of sight extinction but it affects the 
scattering angle; \\
c) the grains are aligned by the macroscopic velocity field of the
Homunculus such that it is their alignment that controls the polarization 
rather than the individual grains. A rather extreme alignment such as
strings of dust particles would be required so it would be the incident
radiation on a grain rather than its intrinsic scattering properties
which would have a greater effect. Given that there are highly collimated 
ejecta observed outside the Homunculus (Weis \& Duschl \cite{weis} and
Morse et al. \cite{morse}), and that such features appear to extend
inwards towards $\eta$ Carinae, the suggestion of an influence of
the macroscopic grain alignment on the scattered radiation may not be
completely ruled out.

  Suggestion (c) finds support in the detection of 12.5$\mu$m polarization
by Aitken et al. (\cite{ait95}), who first showed that there is organized
grain alignment in the Homunculus. The maximum values of 12.5$\mu$m 
polarization were about 5\% and the E-vectors are oriented mostly radially 
at the edges (Aitken et al. \cite{ait95}, Fig. 2), although the
pattern is complex. It is notable that the 12.5$\mu$m polarization is largest
in each lobe where the near-IR polarization is greatest - viz. at the
ends of the lobes. This suggests an intrinsic connection between
the grains at the two wavelengths, rather than the polarization arising
in different grain groups at the different wavelengths.  
Aitken et al. (\cite{ait95}) discuss radiation streaming as
a mechanism to provide suprathermal grain spin of 
paramagnetic grains which then precess about the magnetic field
direction. Fields strengths upto milli-Gauss were suggested and the field
orientation in the lobes orthogonal to the major axis was favoured
(Aitken et al. \cite{ait95}). One possible scenario which could relate
the presence of aligned grains and the constancy of optical - IR 
polarization with wavelength has the grains in the foreground lobes 
of the Homunculus acting as the aligning medium for the scattered 
light from the rearside. By comparison with the case of Galactic grain 
alignment, the amount of extinction to produce 30\% polarization (V band) is 
E$_{B-V}$ $^>_\sim$3.3 (see e.g. Whittet, \cite{whitt}, Fig. 4.2). This is a 
large extinction but not ruled out given the deduced extinction of a few 
magnitudes for the `dark' regions of the Homunculus (see Table 2). In 
addition the grains in the Homunculus are probably very different from 
those in the general interstellar medium, having been recently ejected from a 
star with anomalous abundances. 
It is clear that grain alignment cannot be wholly responsible for the constancy of 
the polarization with wavelength but may be a contributor. Further 
observations to explore the polarization in the 3-5$\mu$m region where
there is a mixture of scattering and emission would be particularly valuable. 
 
\section{Conclusions}
The first high spatial resolution adaptive optics near-IR polarization 
maps of $\eta$ Carinae and the Homunculus nebula have been presented.
Since the Homunculus is dominated by scattering then the appearance
in the near-IR is rather similar to that observed in the optical and a 
comparison of the AO results with an HST 1.04$\mu$m WFPC2 image
was presented showing essentially the same features. The most important 
single result from this work is the overall similarity of the linear 
polarization from the V band to 2.2$\mu$m within a few percent for the 
SE lobe and the lower values at J and H compared with V for the NW lobe. 
Image restoration was applied to a set of 2.15$\mu$m 
continuum images to determine the polarization distribution in the 
near vicinity of $\eta$ Carinae. A tentative value of the polarization of
the Weigelt et al. speckle knot D of 18\% was determined suggesting
that it is a dust cloud within the equatorial disc strongly illuminated 
by $\eta$ Carinae. Various models are discussed in order to explain the 
flat distribution of polarization with wavelength over the Homunculus.
A possible association of a narrow feature within the 
NW lobe of the Homunculus with one of the highly collimated 
emission line `whiskers' outside the nebula deserves further
investigation.

\end{document}